%% file: Ramanv6.tex
\documentclass[a4paper, oneside, twocolumn, notitlepage, 10pt]{extarticle_ecoc2014}
\usepackage{ecoc2014}
\usepackage{amsmath,amssymb}
\usepackage{mathptmx,graphicx}
\input{definitions}
\newcommand{\uu}{\underline}
\newcommand{\onlinecite}[1]{\hspace{-1 ex} \nocite{#1}\citenum{#1}}

\begin{document}
\selectlanguage{american}    


\title{Mitigation of inter-channel nonlinear interference in WDM systems}


\author{
    Ronen Dar\textsuperscript{(1)}, Omri Geller\textsuperscript{(2)}, Meir Feder\textsuperscript{(3)}, Antonio Mecozzi\textsuperscript{(4)}, and Mark Shtaif\textsuperscript{(5)}
}

\maketitle                  


\begin{strip}
 \begin{author_descr}
 
\textsuperscript{(1,2,3,5)} School of Electrical Engineering, Tel Aviv University, Tel Aviv 69978, Israel

\textsuperscript{(4)} Department of Physical and Chemical Sciences, University of L'Aquila, 67100 L'Aquila, Italy
\center{ \textsuperscript{(1)} \uline{ronendar@post.tau.ac.il}}

 \end{author_descr}
\end{strip}

\setstretch{1.1}


\begin{strip}
  \begin{ecoc_abstract}
    We demonstrate mitigation of inter-channel nonlinear interference noise (NLIN) in WDM systems for several amplification schemes. Using a practical decision directed recursive least-squares algorithm, we take advantage of the temporal correlations of NLIN to achieve a notable improvement in system performance. 
  \end{ecoc_abstract}
\end{strip}

\section{Introduction}
\vspace{-0.27cm}

Nonlinear interference noise (NLIN) that is caused by the nonlinear interaction between different channels in a WDM system is a major factor in limiting the capacity of fiber-optic systems \cite{Essiambre}. Often, it is convenient to treat NLIN on the same footing as one treats amplified spontaneous emission (ASE) noise \cite{Essiambre,Ellis,MecozziEssiambre}. However when it comes to system design this approach is clearly sub-optimal as it does not take advantage of the fact that NLIN, unlike ASE, is not white noise, but rather it is characterized by temporal correlations that may be exploited to improve performance.

In this paper our goal is to examine the extent to which the temporal correlations of inter-channel NLIN can be exploited for its mitigation. The underlying idea was reported in [\onlinecite{DarOFC14}], and it is that temporal correlations make NLIN equivalent to a linear inter-symbol-interference (ISI) impairment with slowly varying coefficients, and hence the task of its mitigation can be approached by means of linear equalization with adaptive coefficients. The difference with respect to [\onlinecite{DarOFC14}] is that we are considering a practical equalization algorithm and applying it to a realistic system involving 5 pol-muxed 16-QAM channels, operating at 32 Giga-baud, and spaced by 50 GHz from one another, using lumped or Raman amplification.
The two polarization samples of the $n$-th received symbol in the channel of interest can be expressed as $\uu r_n = \uu a_n + \Delta \uu a_n +\uu z_n$, where $\uu a_n$ is the vector of transmitted data-symbols, $\Delta \uu a_n$ is the NLIN vector, and $\uu z_n$ accounts ASE noise. As in [\onlinecite{DarOFC14}], $\Delta \uu a_n$ is given by
\be  \Delta \uu a_n = \sum_k \mathbf H_k^{(n)}\uu a_{n-k} \ee
where $\mathbf H_k^{(n)}$ is a $2\times2$ interference matrix, which generalizes the scalar ISI coefficient appearing in the single polarization case \cite{DarOFC14}. The superscript $(n)$ in $\mathbf H_k^{(n)}$ accounts for the time dependence of the interference coefficients, whereas we refer to the index $k$ as the ``interference order.'' While in the single polarization case \cite{DarOFC14}, the zeroth interference order was equivalent to phase-noise, in the pol-muxed scenario, only the diagonal terms of $\mathbf H_0^{(n)}$ correspond to phase noise, whereas the off-diagonal terms represent interference between the two polarizations, mediated by the nonlinear interaction with neighboring channels. We show in what follows that the NLIN contributions of the diagonal and off-diagonal terms are similar in magnitude. As we also show in what follows, the zeroth interference order $\mathbf H_0^{(n)}$ is the one whose time dependence is the slowest, and hence its contribution to NLIN is the most amenable to mitigation. We show that the extent to which the system as a whole is amenable to NLIN cancelation strongly depends on the amplification strategy.
In distributed amplification systems the zeroth-order interference term is dominant and its time correlation is the longest. For this reason NLIN mitigation is very effective with this scheme. With lumped amplification the relative significance of the zeroth-order interference term is smaller and also its correlation length is shorted than it is with distributed amplification. Therefore, a system with lumped amplifiers is the least amenable to NLIN mitigation. A practical compromise is a standard Raman amplified system, for example, of the kind demonstrated in [\onlinecite{Chongjin1}]. As we show in this paper, the performance improvement that follows from NLIN mitigation in this case is quite notable.

Due to the very long computation times that are implied by the need to monitor error-rates, we first test our insight regarding time correlations and amplification strategies on a relatively short link, and then simulate a long Raman amplified link so as to characterize the effect on BER.

Finally, as our goal in this work is is to explore the potential of mitigating inter-channel NLIN, we deployed idealized single-channel back-propagation in order to eliminate intra-channel distortions.
\vspace{-0.25cm}
\section{Slowly varying ISI model}
\vspace{-0.05cm}
It can be shown that in the case of a single interfering channel $\mathbf H_k^{(n)} = \tfrac{8}{9}i\gamma\sum_{l,m}(\uu b_{n-m}\uu b_{n-l}^\dagger+\uu b_{n-l}^\dagger\uu b_{n-m}\mathbf I) X_{k,l,m}$, where $\uu b_k$ is the $k$-th vector of data symbols in the interfering channel, $\gamma$ is the usual nonlinearity coefficient and $X_{k,l,m}$ is a coefficient given in [\onlinecite{DarOpex}], and which depends on the transmitted waveform, on the fiber parameters, and on the frequency separation between the channels. In the presence of multiple interferers the matrix $\mathbf H_k^{(n)}$ equals the sum of the matrices representing the contributions of the single interferers.

In the limit of large accumulated dispersion, the number of nonzero terms in the summation becomes very large and the dependence of $\mathbf H_k^{(n)}$ on the time $n$ becomes weak, justifying the formulation in Eq. (1). As it turns out, the zeroth interference matrix $\mathbf H_0^{(n)}$ which produces the strongest interference, is also the slowest changing one, and therefore the one whose estimation from the received data is most accurate. All of the NLIN mitigation examples that are shown in what follows are based on the cancelation of only the $k=0$ term.

\vspace{-0.25cm}
\section{The simulated system}
\vspace{-0.05cm}
We have performed a series of simulations assuming a five-channel polarization multiplexed WDM system implemented over standard single-mode fiber (dispersion of 17 ps/nm/km, nonlinear coefficient $\gamma = 1.3$ [Wkm]$^{-1}$, and attenuation of 0.2dB per km). The transmission consisted of Nyquist pulses at 32 Giga-baud and a standard channel spacing of 50 GHz. The number of simulated symbols in each run and for each polarization was 4096 when simulating a 500 km system, and 8192 when simulating 6000 km. More than 100 runs (each with independent and random data symbols) were performed with each set of system parameters, so as to accumulate sufficient statistics. In the Raman amplified cases, a combination of co-propagating and counter-propagating Raman pumps, providing gains of 5 dB and 15 dB, respectively, as in  [\onlinecite{Chongjin1}].
At the receiver, the channel of interest was isolated with a matched optical filter and back-propagated so as to eliminate the effects of self-phase-modulation and chromatic dispersion. The noise figure of the lumped amplifiers was 4dB and the Raman noise was generated with $n_{sp}=1.26$ (corresponding to a local NF of 4dB).
To cancel the effect of the interference matrices $\mathbf H_k^{(n)}$ we trained and tracked a canceling matrix, using a standard decision-aided approach based on the recursive least squares (RLS) algorithm \cite{RLS}. The RLS algorithm converged faster than the more familiar least mean-squares (LMS) algorithm, and with a small filter order, the added complexity is tolerable. The algorithm uses a ``forgetting factor'' $\lambda$ which weights the past samples exponentially, where a smaller factor allows faster adaptation while a larger factor results in higher stability and convergence. In the experiments we used $\lambda = 0.98, 0.96, 0.9$, specified for the cases of lumped, Raman and distributed amplification, respectively.

\begin{figure}[t]
\center
\epsfig{file=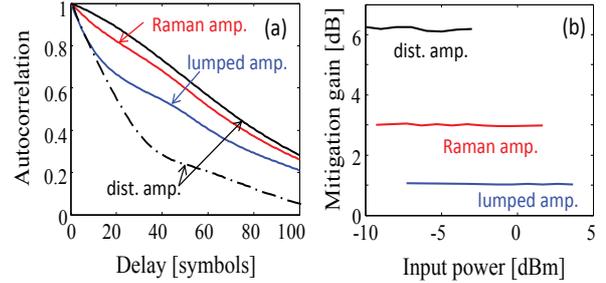,height=0.25\textwidth,width=0.5\textwidth,clip=}
\caption{(a) Solid curves are the temporal autocorrelation functions (ACF) of one of the elements of $\mathbf H_0^{(n)}$ (all elements of $\mathbf H_0^{(n)}$ have practically indistinguishable ACFs). The dash-dotted curve is the ACF of one of the elements of $\mathbf H_1^{(n)}$ in the distributed amplification case. (b) The ratio between the NLIN power before and after mitigation of the zeroth interference term $\mathbf H_0^{(n)}$. In both figures, the colors correspond to lumped (blue), Raman (red), and distributed (black) amplification. Here, unlike in Figs. \ref{SNR} and \ref{BER}, the simulation was performed without ASE noise.}
\label{Corr}
\end{figure}

\vspace{-0.25cm}
\section{Results and discussion}
\begin{figure}[t]
\center
\epsfig{file=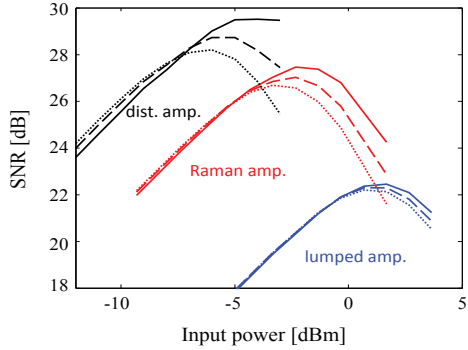,height=0.3\textwidth,width=0.4\textwidth,clip=}
\caption{The effective SNR defined as the signal power divided by the sum of the ASE and NLIN powers in a 500km link. Black, red and blue correspond to distributed, Raman, and lumped amplification, respectively. Without inter-channel NLIN mitigation (dotted), after mitigation of the zeroth NLIN term $\mathbf H_0^{(n)}$ (solid), and after cancelation of only the diagonal elements in $\mathbf H_0^{(n)}$ (dashed). Evidently, the gains in the peak effective SNR are 0.3 dB (lumped), 0.9 dB (Raman) and 1.3 dB (distributed), where approximately half is attributed to the mitigation of the diagonal elements of $\mathbf H_0^{(n)}$. In the linear regime, imperfect estimation of the equalizer coefficients (due to ASE) is seen to slightly spoil performance.}
\label{SNR}
\end{figure}

\vspace{-0.05cm}
To gain insight on the temporal correlations we plot in Fig. \ref{Corr}a the autocorrelation function of one of the diagonal elements of $\mathbf H_0^{(n)}$ in the three amplification strategies after 500km of propagation (the four elements of $\mathbf H_0^{(n)}$ have practically indistinguishable ACFs). As can be seen in the figure, the correlations extend over a few tens of symbols, which is the key factor behind the implementation of the proposed mitigation approach. The autocorrelation of the elements of $\mathbf H_k^{(n)}$ for $k>0$ is generally much shorter, as can be seen from the dash-dotted curve in Fig. \ref{Corr}a showing the ACF of one diagonal element of $\mathbf H_1^{(n)}$ in the distributed amplification case. With the other amplification strategies the ACF of terms with $k\geq 1$ was too narrow to allow its reliable estimation. For this reason, only the mitigation of the $k=0$ matrix is considered in what follows.
In Fig. \ref{Corr}b we show the NLIN mitigation gain defined as the ratio between the NLIN variances before and after  mitigation. As is evident from the figure, the significance of the zeroth interference term $\mathbf H_0^{(n)}$ is largest in the case of distributed amplification and smallest in the lumped amplification case.
In Fig. \ref{SNR} we show the effective SNR curves with (solid) and without (dotted) NLIN mitigation.  The peak SNR after mitigation is larger than the peak SNR prior to mitigation by 0.3 dB (with lumped amp.), 0.9 dB (with Raman amp.) and 1.3 dB (with distributed amp). The dashed curves in the figure represent the case in which mitigation addresses only the diagonal terms of $\mathbf H_0^{(n)}$, showing that the advantage of accounting for NLIN induced interference between polarizations is notable.

\begin{figure}[t]
\center
\epsfig{file=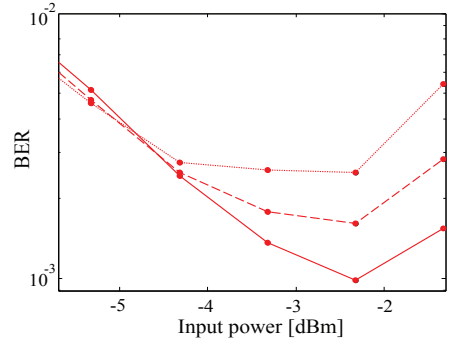,height=0.3\textwidth,width=0.4\textwidth,clip=}
\caption{The BER as a function of input power in a 6000 km Raman amplified system. Dotted -- without mitigating inter-channel NLIN, solid -- mitigation of $\mathbf H_0^{(n)}$, dashed -- mitigation of only the diagonal elements in $\mathbf H_0^{(n)}$. At very low powers mitigation spoils performance as a result of poor estimation caused by ASE.}
\label{BER}
\end{figure}

Finally we address in Fig. \ref{BER} the consequences of mitigation in terms of the system BER. Since truly distributed amplification is of limited practical interest, we chose to concentrate this highly computationally intensive set of simulations on the case of the practical Raman-amplified scenario, where the gain of mitigation is higher than it is in the lumped amplification case (see Fig. \ref{SNR}). The simulations in this case were conducted for $60\times100$ km spans, so as to reach relevant BER values. The dotted curve represents the case without mitigation and the solid curve represents the results obtained after canceling the effect of the zeroth interference term $\mathbf H_0^{(n)}$, showing an improvement of the BER from $2.6\times 10^{-3}$ to less than $1\times 10^{-3}$, equivalent to approximately 1dB gain in the Q factor. Equivalently, to achieve the same BER without NLIN mitigation the OSNR would have to be increased by 2.3dB. Similarly to Fig. \ref{SNR}, the dashed curve shows the result obtained when only the effect of the diagonal terms in $\mathbf H_0^{(n)}$ was mitigated, showing that approximately half of the mitigation gain is attributed to cross polarization interference.

\vspace{-0.25cm}
\section{Conclusions}
\vspace{-0.05cm}
We showed that significant improvement in system performance can be achieved by using adaptive linear equalization methods for mitigating inter-channel NLIN. The proposed scheme has the advantage of using the same type of hardware currently used for equalizing polarization effects, although the equalization algorithm and the speed of convergence are substantially different.

\bibliographystyle{abbrv}
\begin{spacing}{1.15}

\end{spacing}
\vspace{-4mm}

\end{document}

%% file: definitions.tex
\usepackage{caption,subcaption,colordvi,psfrag}
\usepackage{epstopdf}
\usepackage{calc,pstricks, pgf, xcolor}
\usepackage{epsfig}
\usepackage{setspace}
\newcommand{\be}{\begin{equation}}
\newcommand{\ee}{\end{equation}}
\newcommand{\bea}{\begin{eqnarray}}
\newcommand{\eea}{\end{eqnarray}}